\documentclass[aps,pra,twocolumn,superscriptaddress,%
     amsmath,amssymb,showpacs]{revtex4}
\usepackage{psfig}
\usepackage{graphics}
\usepackage{amsmath}

\newcommand{\bra}[1]{\langle#1|}
\newcommand{\ket}[1]{|#1\rangle}
\newcommand{\be}{\begin{equation}}
\newcommand{\ee}{\end{equation}}
\newcommand{\bea}{\begin{eqnarray}}
\newcommand{\eea}{\end{eqnarray}}
\newcommand{\bsube}{\begin{subequations}}
\newcommand{\esube}{\end{subequations}}
\newcommand{\Fig}[1]{Fig.\,\ref{#1}}
\newcommand{\Eq}[1]{Eq.\,(\ref{#1})}

\begin{document}

\title{Transverse localization and slow propagation of light
}

\author{Jing Cheng}

\affiliation{Department of Chemistry, Hong Kong University 
of Science and Technology, Kowloon, Hong Kong}

\affiliation{Key Laboratory for Quantum Optics 
and Center for Cold Atom Physics, Shanghai Institute of 
Optics and Fine Mechanics, 
Chinese Academy of Sciences, 
Shanghai 201800, China} 

\author{Shensheng Han}
\affiliation{Key Laboratory for Quantum Optics
and Center for Cold Atom Physics, Shanghai Institute of
Optics and Fine Mechanics, 
Chinese Academy of Sciences, 
Shanghai 201800, China}

\author{YiJing Yan}
\affiliation{Department of Chemistry, Hong Kong University 
of Science and Technology, Kowloon, Hong Kong}

\begin{abstract}
 The effect of finite control beam on the
transverse spatial profile of the slow light propagation in 
an electromagnetically induced transparency medium is studied. 
We arrive at a general criterion in terms of eigenequation, 
and demonstrate the existence of a set of localized, stationary 
transverse modes for the negative detuning of the probe signal field.
Each of these diffraction-free transverse modes 
has its own characteristic group velocity,
smaller than the conventional theoretical result
without considering the transverse spatial effect.
\end{abstract}

\pacs{42.50.Gy, 42.65.-k}

\maketitle

Ultraslow propagation of light fields 
has been an active research field recently. 
Controlling the light propagation 
in atomic and solid-state media is important  
in both the fundamental theory and practical
applications of nonlinear optics \cite{rev}. 
Slow light propagation experiments have been reported by using 
ultracold atoms \cite{Hau99,Inouye00,kane04}, hot atoms 
\cite{Kash99,Budker99}, rare-earth ion doped crystal \cite{turukhin02},
ruby \cite{bigelow03a}, and alexandrite crystals \cite{bigelow03b}.
The use of electromagnetically induced transparency (EIT) to 
obtain slow light propagation is one of the most important techniques. 
In an EIT medium \cite{eit}, 
when a control laser is applied to an appropriate transition, 
a weak probe signal pulse may have small absorption and steep dispersion.
Due to this steep dispersion, the group velocity of the signal pulse 
can be reduced to several orders smaller than the light speed in 
vacuum \cite{Hau99,Inouye00,kane04,Kash99,Budker99,turukhin02}.
Further, by changing the intensity of the control laser, it is possible
to reversibly stop the signal pulse \cite{polarit}. 
Stop light propagation has been observed in cold and 
hot alkali vapors \cite{liu01,phillips01}.
Possible applications of slow light include 
the enhancement of nonlinearity
\cite{Schmidt96,Harris98,Masalas04}, entanglement of atomic ensembles 
or photons \cite{Lukin00a,Lukin00b}, quantum memories \cite{qmemo}, 
and optical information processing \cite{infbec}.

Almost all theoretical treatments so far
involve the assumption of an effective infinite transverse
spatial variation of control field, 
and how a finite transverse profile
affects the light propagation in an EIT medium
is yet to be addressed.  
An exception is the work in Ref.\ \onlinecite{Andre04}. 
The authors investigated the transverse localization of the 
stationary probe pulses with a pair of counter-propagating control 
fields, and showed  
it is possible to realize a three-dimensional confinement of the 
probe pulses \cite{Andre04}. The focus of Ref.\ \onlinecite{Andre04} 
was how to localize light pulses in three dimensions. 
In this paper, we are interested in the
transverse effects on slow light propagation. 
In particular, we arrive at a general criterion in terms of eigenequation 
for the transverse stationary modes of the signal field. 
These transverse modes exist only at negative detuning frequency region, and 
do not undergo diffraction, thus keeping their profiles during the 
propagation. The group velocity of each stationary 
mode will also be studied quantitatively.

\begin{figure}[htb]
\centerline{
\begin{tabular}{cc}
\psfig{file=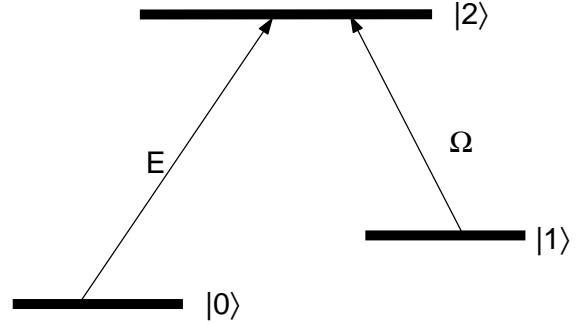,width=0.9\linewidth}
\end{tabular}
}
\caption{ The three-level $\Lambda$-type EIT medium. 
$\Omega$ is the Rabi frequency of the classical control field,
and $E$ is the electric field of the probe signal pulse.
}
\label{model}
\end{figure}

We start with the three-level $\Lambda$-type atoms shown in 
\Fig{model}. The medium of length $L$ consists of an ensemble 
of $N$ atoms. 
The ground state $\vert 0\rangle$ and the metastable Stokes state $\vert 1 
\rangle$ 
are coupled individually with the excited state $\vert 2\rangle$ via a weak  
probe 
signal pulse and an intense control laser field, respectively.
The latter is treated classically and assumes
the form $\mathcal{E}_c(z,t,\vec{r})=\Omega(z,t,\vec{r})e^{i(k_c z -i\omega_c  
t)}$, 
where $\Omega(z,t,\vec{r})$ denotes the Rabi frequency and $k_c=\omega_c/c$,
with the carrier frequency, $\omega_c=\omega_{12}$,
being set to be on resonance with the Stokes transition.
The weak signal field shall be treated as a quantum field \cite{polarit}:
\be \label{eq1}
\hat{{\mathcal{E}}}(z,t,\vec{r})
=\hat{E}(z,t,\vec{r})e^{i(k_0 z -i\omega_{02} t)},
\ee
where $k_0\equiv \omega_{02}/c$.
The probe carrier frequency $\omega_s$ is assumed to be close to 
the $\vert 0\rangle \to \vert 2\rangle$ transition frequency 
$\omega_{02}$.
In \Eq{eq1},
$\hat{E}(z,t,\vec{r})$ denotes the slowly varying 
signal field envelop operator.
Following Ref.\ \onlinecite{polarit}, but including now also the 
finite control beam effects, the signal pulse 
in the paraxial approximation and slowing varying amplitude approximation
can be describe by the propagation equation,
\be 
\left(\frac{\partial}{\partial t}+ c\frac{\partial}{\partial z}
-i\frac{c\nabla_{\rm T}^2}{2k_0}\right)\hat{E}
=igN\hat{\sigma}_{02}(z,t,\vec{r}) .
\label{propE}
\ee
Here, $\nabla_{\rm T}^2=\nabla^2 - \frac{\partial^2}{\partial z^2}$
is the transverse Laplacian, 
 $g=\mu_{02}\sqrt{\frac{\omega_{02}}{2\hbar\epsilon_0 V}}$;
with $\mu_{02}$ being the transition dipole moment and $V$ the quantization  
volume,
denotes the atom-field coupling constant for $\vert 0\rangle \to \vert 2 
\rangle$,
while $\hat{\sigma}_{02}$ is a slowly varying 
collective operator of the atoms. In general, 
$\hat \sigma_{ij}(z,t,\vec{r})=
\frac{1}{N_{\vec{r}}}
\sum_{l=1}^{N_{z,\vec{r}}}\ket{i}_l\bra{j}e^{-i\omega_{ij}t}$,
where the sum runs over the effective number of atoms 
in a small but macroscopic volume around position $\vec{r}$ 
\cite{Andre04}. 

  We shall be interested in the case in which
the Rabi frequency of the signal field is much 
smaller than that of the control field $\Omega$
and the number of input probe photons 
is much less than that of atoms.
In the adiabatic approximation, 
$\hat{\sigma}_{02}(z,t,\vec{r})=\frac{ig}{\Omega}
  \frac{\partial}{\partial t}\frac{\hat{E}(z,t,\vec{r})}{\Omega}$
\cite{polarit,Andre04},
leading \Eq{propE} to 
\be
\left(\frac{\partial}{\partial t}+ c\frac{\partial}{\partial z}
-i\frac{c\nabla_{\rm T}^2}{2k_0}\right)\hat{E}
=-\frac{g^2N}{\Omega}
  \frac{\partial}{\partial t}\frac{\hat{E}(z,t,\vec{r})}{\Omega}
\label{mainEq}
\ee
In most experiments, the control field is continuous
and has a cylindrical symmetry transverse spatial profile $\Omega(r)$
that changes little in the propagation direction. 
To study the transverse effects of the probe signal field
in this case, we consider the expectation value of $\hat{E}(z,t,\vec{r})$ 
in terms of
\be
E(z,t,\vec{r})=\psi(r)e^{im\theta}e^{i(\Delta z-\delta t)} ,
\label{Eform}
\ee
with the quantum number $m=0,\pm 1, \cdots$ for
the orbital angular momentum of the signal field \cite{OAM}.
The signal wavevector mismatch $\Delta\equiv k_s - k_0$ 
along the $z$-direction
will be determined as the function
of frequency detuning $\delta \equiv \omega_s- \omega_{02}$.
Equation (\ref{mainEq}) can then be reduced to 
\be
 \left[-\frac{d^2}{dr^2}-\frac{1}{r}\frac{d}{dr}
-\frac{2k_0g^2N\delta}{c\Omega^2(r)} +\frac{m^2}{r^2} 
\right]\psi(r)=\beta\psi(r),
\label{eigenEq}
\ee
where the eigenvalue $\beta\equiv 2k_0(\delta/c-\Delta)$
defines the dispersion relation between $\Delta$ 
and $\delta$ for each transverse mode of the signal field.
The physical boundary conditions 
for \Eq{eigenEq} are 
$\psi(\infty)=\psi'(0)|_{m=0}=\psi(0)|_{m\neq 0} =0$.

  It is noted that \Eq{mainEq} may be considered as a 
propagation version of Eq.(2) in Ref.\ \onlinecite{Andre04}
(by setting $\Omega_{-}=0$ there). 
However, our work and Ref.\ \onlinecite{Andre04} have different physical 
background. Ref.\ \onlinecite{Andre04} was focused on how to realize 
light localization. 
The novel result of this paper is \Eq{eigenEq}. It unambiguously 
shows the general properties of a complete set of 
transverse invariant eigenmodes for slow light propagation in the EIT medium. 

For a negative detuning ($\delta<0$), 
the eigenequation (\ref{eigenEq}) for each integer value of $m$ is quantized.
The resulting transverse modes and eigenvalues are denoted
as $\psi_{mn}(r)$ and $\beta_{mn}$, respectively;
with $n=1, 2, \cdots$ and $0<\beta_{m1}<\beta_{m2}< \cdots$
for each $m=0, \pm 1, \cdots, \pm (n-1)$.
Physically, the negative frequency detuning 
leads to the decrease of the refractive index from the optical axis 
and the production of an effective waveguide for the signal field.
The group velocity for each stationary transverse mode
of the signal field can be evaluated via its eigenvalue as
\be
 V_g^{mn}=\left(\frac{\partial \Delta}{\partial \delta}\right)^{-1}
   =\left(1-\frac{c}{2k_0}\frac{\partial \beta_{mn}}{\partial \delta}
              \right)^{-1}c .
\label{Vg}
\ee
For a given control field with an arbitrary transverse profile
and the relevant parameters for the optical medium,
one can solve \Eq{eigenEq} to obtain all eigenvalues $\beta_{mn}$ and 
the corresponding transverse modes $\psi_{mn}(r)$. 
Any localized input probe signal pulse $E(0,t,\vec{r})$ can be 
expanded as the superposition of these transverse modes;
each of these modes then propagates at its own distinct velocity,
and thus, the signal pulse of superposition 
changes its spatial-temporal profile during the 
propagation. Clearly, if the input pulse is characterized by a 
single transverse mode it will not undergo diffraction and 
remains the initial shape during the propagation.

In the following we use numerical simulations to 
demonstrate the effects of control light beam size
on various stationary transverse signal modes 
with negative detuning. 
The transverse spatial profile of the 
control field is chosen to be Gaussian,
\be
  \Omega(r) = \Omega_0 e^{-r^2/(2a^2)} .
\ee 
The reported results will be exemplified with
$\Omega_0=10^8\,{\rm s}^{-1}$ and $a=50\,\mu{\rm m}$. 
The parameters of the medium 
are set to be $g^2N=10^{22}{\rm s}^{-2}$ for the atom 
density of $10^{14}\,{\rm cm}^{-3}$,
and the $\vert 0\rangle \to \vert 2\rangle$ transition wavelength is 780\,nm.
The eigenmodes are normalized as
$
\int_{0}^{\infty} r|\psi_{mn}(r)|^2dr={\rm const}
$.

\begin{figure}[htb]
\centerline{
\begin{tabular}{cc}
\psfig{file=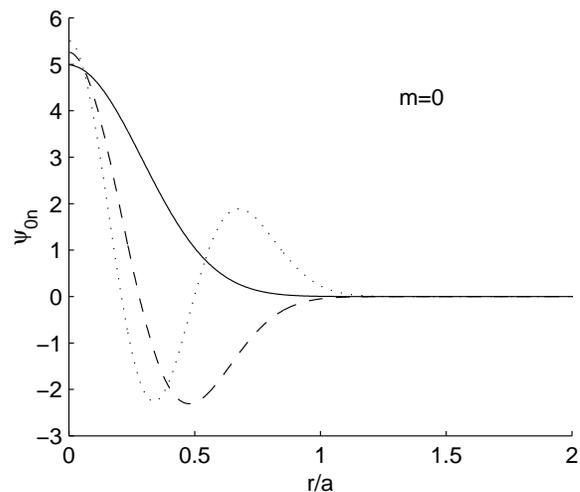,width=0.9\linewidth}
\end{tabular}
}
\caption{The lowest three transverse modes for $m=0$:
$\psi_{01}(r)$ (solid),  
$\psi_{02}(r)$ (dash), 
and $\psi_{03}(r)$ (dot);
with the probe signal detuning $\delta=-10^6\,{\rm s}^{-1}$.
See text for the other parameters.
}
\label{exm1f01}
\end{figure}

\begin{figure}[htb]
\centerline{
\begin{tabular}{cc}
\psfig{file=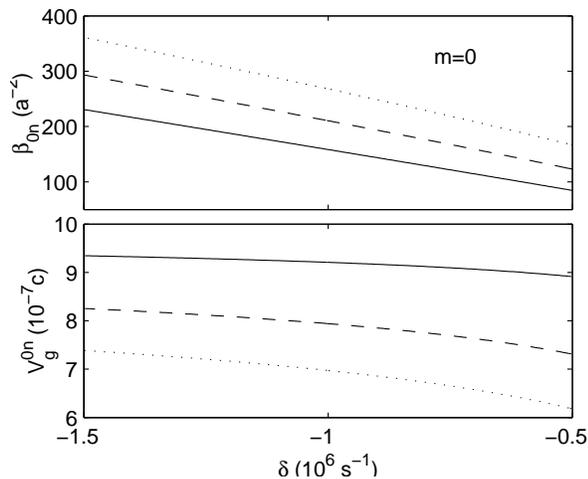,width=0.9\linewidth}
\end{tabular}
}
\caption{The eigenvalues (upper panel) and the 
corresponding group velocities (lower panel),
for the transverse eigenmodes $\psi_{01}(r)$ (solid lines),
$\psi_{02}(r)$ (dashed lines), and $\psi_{03}(r)$ (dotted lines),
as the functions of detuning $\delta$.
Other parameters are the same as in \Fig{exm1f01}.
}
\label{exm1Vg}
\end{figure}

Figure \ref{exm1f01} shows the radial profiles of 
the lowest three transverse modes, 
$\psi_{01}(r)$, $\psi_{02}(r)$, and $\psi_{03}(r)$;
with $m=0$ and $\delta=-10^6\,{\rm s}^{-1}$. 
The lowest mode $\psi_{01}(r)$ decreases monotonically,
while the higher mode $\psi_{0n}(r)$, with $n>1$,
oscillates and has $(n-1)$ nodes. 
Here, we must point out that numerical calculations show that 
Gaussian function can well approximate the 
the ground mode ($\psi_{01}(r)$) \cite{Andre04}.
Figure \ref{exm1Vg} depicts the eigenvalues $\beta$ (upper panel)
and the corresponding group velocities $V_g$ (lower panel)
of the lowest three transverse modes 
as functions of detuning $\delta<0$.
For a given transverse mode,
an increase in $|\delta|$ leads to an increase
in $|\beta|$ (and also in wavevector mismatch),
but a decrease in the group velocity $V_g$ 
as it is calculated according to \Eq{Vg}.
Note that if the effect of transverse spatial distribution is
completely neglected, the group velocity for the present system would be 
$V_g =(1+g^2N/\Omega_0^2)^{-1}c=10^{-6}c$, 
larger than the values we obtained here. 
It is also interesting to see that at a given negative detuning,
a higher mode is of a smaller group velocity. 
Thus, by making use of a high-order transverse mode of 
the probe signal field with a small negative detuning frequency,
it is possible to have a slow light propagation in
an EIT medium. 

\begin{figure}[htb]
\centerline{
\begin{tabular}{cc}
\psfig{file=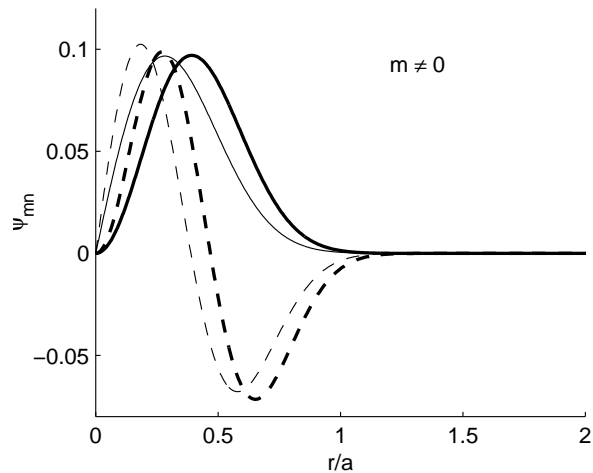,width=0.9\linewidth}
\end{tabular}
}
\caption{Some lowest transverse modes with finite orbital 
angular momentum $m$: $\psi_{11}(r)$ (thin-solid),
$\psi_{12}(r)$ (thin-dash),
$\psi_{21}(r)$ (thick-solid),
and $\psi_{22}(r)$ (thick-dash).
Other parameters are the same as in \Fig{exm1f01}.
}
\label{exm1fm}
\end{figure}

\begin{figure}[htb]
\centerline{
\begin{tabular}{cc}
\psfig{file=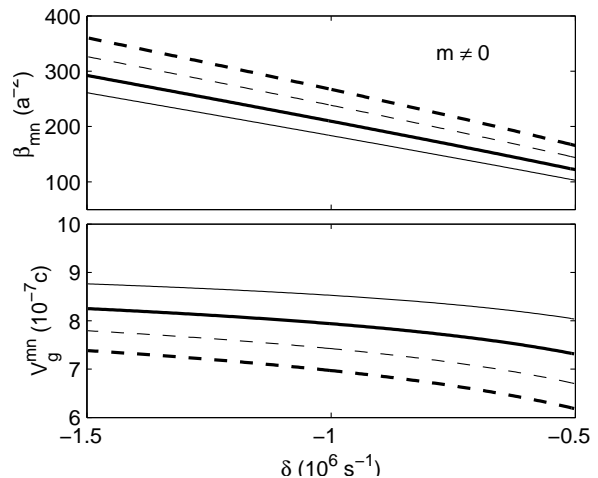,width=0.9\linewidth}
\end{tabular}
}
\caption{The eigenvalues (upper panel) and the group velocities (lower paner)
for the eigenmodes of \Fig{exm1fm} as the functions
of detuning frequency $\delta$.
}
\label{exm1mVg}
\end{figure}

  We now study the transverse modes with nonzero orbital angular 
momentum, i.e., the eigenmodes of \Eq{eigenEq} 
with $m\neq 0$, which represent optical vortex. 
Note that $\psi_{-m,n}=\psi_{m,n}$ [c.f.~\Eq{eigenEq}].
In \Fig{exm1fm}, we plot the ``ground'' and the first ``excited'' 
modes for both $m=1$ and $m=2$: $\psi_{11}$ (thin-solid),
$\psi_{12}$ (thin-dash), $\psi_{21}$ (thick-solid), 
and $\psi_{22}$ (thick-dash), 
for the same EIT system and field parameters of \Fig{exm1f01}.
The transverse mode $\psi_{mn}(r)$ is found to have $(n-1)$ nodes,
besides that of $\psi_{m\neq 0,n}(0)=0$.
For a given $n$, $\psi_{mn}(r)$ with a large $m$ 
extends to a large $r$.
The negative detuning frequency dependences of 
these $m\neq 0$ modes are presented 
in \Fig{exm1mVg}. The qualitative properties are similar as 
\Fig{exm1Vg}. 
It is found that $V_{g}^{11}>V_g^{21}> V_g^{12} > V_g^{22}$.
The group velocities for
the transverse mode $\psi_{mn}$ 
satisfy in general
\be
V_g^{m1}>V_g^{m2}> \cdots \ \ {\rm and} \ \
V_g^{1n}>V_g^{2n}> \cdots
\ee
at a given negative detuning, 
while each individual $V_g^{mn}(\delta)$ decreases as the negative detuning  
reduces.

All these calculated transverse mode velocities are smaller than
the value of $10^{-6}c$, the group velocity with no
consideration of the finite transverse distribution.
Our calculations also conclude that
the more focused (smaller $a$) the control field is,
the smaller group velocity will be.
On the other hand, as the size of the control field 
increases, the group velocity of each transverse 
eigenmode approaches to the 
asymptotic, transverse-effect-free 
value of $V_g=(1+g^2 N/\Omega_0^2)^{-1}c$,
which is $10^{-6}c$ in the present
EIT system of study.
It also suggests that in order to observe
the transverse effects experimentally,
should not only the propagation length be long enough,
but also the control field be well focused.

 It is noticed that 
the transverse profile of the control field,
$\Omega(r)$  may vary in a realistic propagation in 
the $z$-direction, but it  
is assumed to be stationary in our theoretical treatment. 
The justification here is the fact that
the adiabatic approximation is applicable
if $\Omega(r)$ is a slowly varying function 
of the propagating distance.
As a result, the probe signal field prepared initially in an 
eigenmode $\psi_{m,n}(r)$ can stay in this mode at 
any distance $z$, as long as the adiabatic approximation is applicable.
For a Gaussian control beam, the adiabatic 
approximation requires the propagating distance smaller than the 
Rayleigh length of the control beam. In our example, 
the Gaussian beam with 
a spread of $a=50\,\mu{\rm m}$ leads to 
a Rayleigh length of $10\,{\rm cm}$.
Both the beam focusing size and propagation length parameters
demonstrated here
are well accessible in current experiments.
Moreover, the stationary behavior of an
eigenmode propagation of the signal field can 
sustain over a much longer distance,
if a non-diffracting Bessel rather than 
a Gaussian control beam is used \cite{nondiff}. 

  In conclusion, we have studied the transverse effects of the slow 
light propagation in EIT mediums. In terms of eigenequation, a criterion 
of localized transverse modes is given. Using this criterion,  
a complete set of transverse modes can be numerically calculated. 
Different transverse modes propagate with different group velocities;
all are smaller than the limiting value calculated from the 
previous theory without the consideration of the 
transverse spatial effect.
Increasing the size of the control field will decrease this velocity 
difference. Higher order, or larger orbital angular momentum mode will 
have smaller velocity. 
Our results will also play important roles in 
the coherent light propagation control in general.
For example, in Ref.\ \onlinecite{Andre04}, a
transverse light guiding technique
was used to obtain three-dimensional confinement of
light pulses. A Gaussian approximation to the lowest transverse 
mode with zero orbital angular momentum ($\psi_{01}$) was given. 
With the method present in this work,
we should be able to identify the relevant transverse eigenmodes 
there. By carefully manipulating the transverse
profile of the signal pulse, generalized and efficient
realizations of the controlled location and storage 
of photonic pulses will become possible.

Support from the National Natural Science 
Foundation of China (10404031), 
Shanghai Rising-Star Program,
the K.~C.~Wong Education Foundation (Hong Kong), 
and the Research Grants Council of the Hong Kong Government (604804) 
is acknowledged.

\end{document}